\documentstyle[aps,multicol,epsf]{revtex}
\begin{document}
\draft
\title
{Parity-broken ground state for the spin-1 pyrochlore antiferromagnet}
\author{Yasufumi Yamashita$^1$, Kazuo Ueda$^{1,2}$ and Manfred Sigrist$^{3,4}$}
\address{
$^1$Institute for Solid State Physics, University of Tokyo, 
5-1-5 Kashiwa-no-ha, Kashiwa, Chiba 277-8581, Japan}
\address{
$^2$Advanced Science Research Center, Japan Atomic Energy Research Institute, Tokai, Ibaraki 319-1195, Japan
}
\address{$^3$ Yukawa Institute for Theoretical Physics, Kyoto
  University, Kyoto 606-8502, Japan}
\address{
$^4$Theoretische Physik , ETH-H\"onggerberg CH-8093 Z\"urich, Switzerland
}
\date{Received \today}
\maketitle
\begin{abstract}
The ground-state properties of the spin-1 pyrochlore antiferromagnet 
are studied by applying the VBS-like tetrahedron-unit decomposition 
to the original spin system.
The symmetrization required on every vertex is taken into account by 
introducing a ferromagnetic coupling.
The pairwise effective Hamiltonian between the adjacent tetrahedrons 
is obtained by considering the next nearest neighbor and the third 
neighbor exchange interactions.
We find that the transverse component of the spin chirality exhibits 
a long-range order, breaking the parity symmetry of the tetrahedral group,
while the chirality itself is not broken.
\end{abstract}
\pacs{PACS numbers: 75.10.Jm, 75.40.Cx, 75.10.-b}
\begin{multicols}{2}
\narrowtext
The pyrochlore lattice, the network of the corner sharing tetrahedrons
or a fcc-array of tetrahedrons, is a typical example of a 
three-dimensional(3D) frustrated system, which is found in a number of 
materials, such as spinels, pyrochlores and C15-type Laves phase.
To study effects of the geometric frustrations and 
the resultant enhanced spin fluctuations, the pyrochlore spin systems 
have been investigated intensively. For the spin-1/2 
antiferromagnetic(AF) Heisenberg model, 
the quantum spin-liquid ground state is proposed based on the series
expansion at finite temperature\cite{CANALS}, but concerning the 
properties of the low-lying spin-singlet states below the spin gap, 
a consistent picture has not 
emerged yet\cite{HARRIS,ISODA,KOGA,TSUNETSUGU}.

For the spin-1 case, numerical methods powerful for the spin-1/2 systems
become difficult to obtain definitive results\cite{KOGA} and we need 
analytic approaches to solve the problem. 
We proposed the idea of the tetrahedron-unit decomposition 
of the pyrochlore lattice, which is a natural generalization 
of the valence-bond-solid(VBS) state approach developed for the 
1D spin-1 systems\cite{YAMASHITA}.
In this approach the ground states of the fundamental unit, 
corresponding to the valence bond for the VBS case,
is the tetrahedron spin singlets which form 
the two-dimensional $E$ representation of the tetrahedral group ($T_d$).
In consequence, the constructed VBS-type wavefunctions define the 
variational spin-singlets manifold with a macroscopic degeneracy,
to which the ground state of the original problem may be continued 
adiabatically. In this situation, it is essential to investigate
how the ground state in the thermodynamic limit is stabilized from 
a collection of spin singlets by lifting the degeneracy.

In the previous paper\cite{YAMASHITA}, 
we examined the magneto-elastic couplings between 
the tetrahedron-singlets and the local lattice distortion of the $E$ modes
on the same tetrahedron as the source of lifting the degeneracy.
As a result we have found that a Jahn-Teller mechanism driven by the 
non-magnetic spin degrees of freedom gives rise to the 
structural phase transition. The local lattice distortion is
given by $Q_v$ mode of $T_d$.
Actually the uniform $Q_v$ distortion compressed along the $c$-axis 
is consistent with the cubic to tetragonal structural phase transition 
without accompanying any magnetic order, observed in the spin-1 
spinel-type antiferromagnets, ZnV$_2$O$_4$ and MgV$_2$O$_4$. 
This scenario can explain, at least qualitatively, the experimental results
around the structural transition temperature.

As another way of lifting the degeneracy, in this letter,
we study the case where the inter-tetrahedron interactions,
the next nearest neighbor and the third neighbor spin-spin interactions,
are relevant. In particular, we will discuss the properties of the order 
parameters of the obtained ground state.
Since the tetrahedron-singlets may be labeled by the spin chirality,
a chiral ordered state is an interesting possibility.
However, as long as physically reasonable interactions, such as spin 
exchange interaction or the dipole interaction, are considered, 
we find that it is not the chirality but its transverse component
that shows the long-range order.

\begin{figure}[htb]\begin{center}
\leavevmode
\epsfxsize=86mm
\epsffile{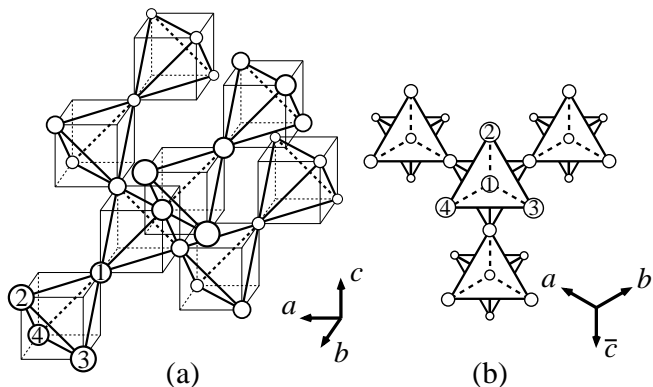}
\caption{
The primitive cell of the pyrochlore lattice (a) and 
its projected view from the [11$\bar{1}$]-axis (b). 
The small cubes in figure (a) are the guide for eyes 
with one-sixteenth volume of the unit cell.
}
\label{FIG1}
\end{center}\end{figure}

Following the 1D VBS approach, 
we break up the original spin-1 objects (denoted by $\vec{S}_i$) 
into the two spin-1/2 ones ($\vec{s}_{ia}$ and $\vec{s}_{ib}$)
and introduce a ferromagnetic Hund couplings between them ($-J_F$),
which serve as the symmetrization. To project out the singlet sector
of the composite spins completely it is necessary to take the 
$J_F\rightarrow \infty $ limit, however in nature a 
strong but finite $J_F$ is sufficient.
The pyrochlore lattice may be decomposed into the $A$ and $B$ types of bonds,
which form the upward and downward tetrahedrons as shown in Fig. \ref{FIG1}(b).
Because of the symmetrization it is allowed to consider that 
the 1/2-spins with $a$ $(b)$ index construct the network of the $A$ $(B)$ 
bonds without loss of generality.
In this decomposite-spin representation, the original AF Heisenberg model can
be rewritten as follows in the strong ferromagnetic coupling limit
$(J_F\rightarrow \infty)$;
\begin{eqnarray}
 {\cal H}_{dec.}&=&4J\sum_{<i,j>\in A}\vec{s}_{ia}\cdot\vec{s}_{ja}
+4J\sum_{<i,j>\in B}\vec{s}_{ib}\cdot\vec{s}_{jb}\nonumber\\
&-&J_F\sum_{i}\vec{s}_{ia}\cdot\vec{s}_{ib}.\label{Hdec}
\end{eqnarray}

\begin{figure}[htb]\begin{center}
\leavevmode
\epsfxsize=86mm
\epsffile{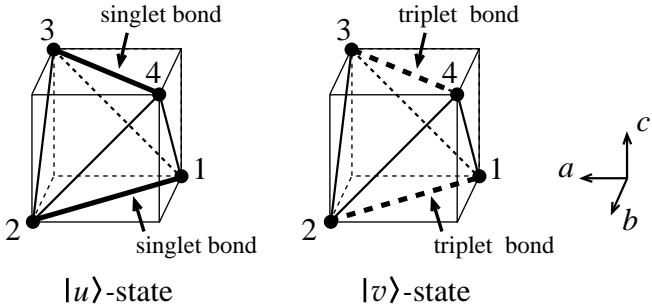}
\caption{
The schematic representations of the tetrahedron-singlet states, 
$|u\rangle$ and $|v\rangle$.
}
\label{FIG2}
\end{center}\end{figure}

In the $J_F=0$ limit, the ground states of Eq. (\ref{Hdec}) are given by;
\begin{eqnarray}
|\Psi _{0}\rangle =\prod_{k=1}^{N/2}
\left(\alpha_k|u\rangle _k+\beta _k |v\rangle _k\right),\label{gs0}
\end{eqnarray}
with arbitrary complex parameters $\alpha_k$ and $\beta _k$ keeping 
$|\alpha _k|^2+|\beta _k|^2 =1$, where $N$ is the number of sites
and $k$ specifies a tetrahedron.
Here the orthonormal basis, $\{|u\rangle,|v\rangle\}$
defined for a single tetrahedron, are the total spin-singlet states
with $\vec{s}_{1x}+\vec{s}_{2x}=\vec{s}_{3x}+\vec{s}_{4x}=0$ or 1, 
respectively, $(x=a,b)$ (see Fig. \ref{FIG2}).
These tetrahedron singlets are the non-magnetic doublets belonging to the 
$E$ representation of $T_d$ and the $2^{N/2}$-fold degenerate manifold 
defined by Eq.(\ref{gs0}) is expected to be adiabatically continued to 
the low-energy sector of the original model.
When representing the symmetry operation of $T_d$
in this singlet subspace, $\{|u\rangle,|v\rangle\}$ real-basis
diagonalizes the parity operations with respect to the bonds vertical
to the $c$-axis. On the other hand, the chirality basis\cite{YAMASHITA}, 
which is defined by
$|R\rangle=\left(|u\rangle-i|v\rangle\right)/\sqrt{2}$ and 
$|L\rangle=|R\rangle ^{*}$, diagonalizes the four distinct $C_3$ operations
with the eigenvalues of $\omega=\left(-1+\sqrt{3}i\right)/2$ and 
its complex conjugate $\omega ^{*}$, respectively.

We proceed to the next step of lifting the $2^{N/2}$-fold degeneracy
of Eq. (\ref{gs0}). 
When considering the pure two-tetrahedron problem, we can easily see
that the ferromagnetic coupling $J_F$ alone does not lift the degeneracy.
Therefore the pairwise perturbation between the two adjacent tetrahedrons 
caused by $J_F$ never fixes the original $2\times 2$-fold degeneracy.
>From the viewpoint of geometry, this means that the relative rotation 
of two tetrahedrons is not fixed by only $J_F$. 
The degeneracy is partly lifted by considering three tetrahedrons
and eventually the degeneracy will be lifted for the entire lattice. 
However energy scales of the lifting of degeneracy are expected to be small
and hierarchical\cite{TSUNETSUGU}. This problem is interesting
but more academic since in nature there are other couplings to lift the
degeneracy. In \cite{YAMASHITA} we have considered the local coupling
with lattice distortion which leads to the structural phase transition.
As another source of lifting the degeneracy we can introduce
longer range interactions which are relevant to lift the degeneracy
already for a pair of tetrahedrons.

\begin{figure}[htb]\begin{center}
\leavevmode
\epsfxsize=86mm
\epsffile{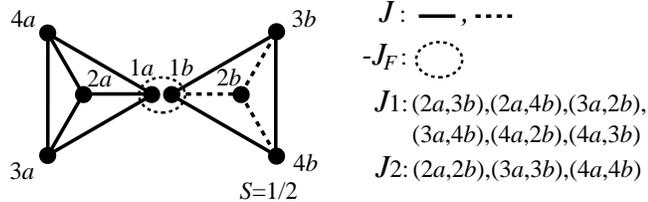}
\caption{
The decomposite-spin representation of the Heisenberg model 
($J$ and infinite $J_F$)
and the next-nearest ($J_1$) and third-neighboring interactions ($J_2$)
are shown for the pairwise ($k_a,k_b$) tetrahedrons.
}
\label{FIG3}
\end{center}\end{figure}

In order to fix the relative rotation of the two adjacent tetrahedrons,
we include the next nearest neighbor ($J_1$) and the third neighbor 
interactions ($J_2$). In spinels, $J_2$ is expected to be important 
because of the existence of a superexchange-path through the oxygens 
on a single plane. The arrangement of these interactions are shown in
Fig. \ref{FIG3}. 
By the second-order perturbation in $-J_F$, $J_1$, and $J_2$ 
within the degenerate subspace spanned by Eq.({\ref{gs0}}), 
we obtain the pairwise effective Hamiltonian between 
the tetrahedron-singlet states on the adjacent tetrahedrons as follows;
\begin{eqnarray}
{\cal H}_{eff}=\frac{1}{2}&&\left(J_2-J_1\right)\left(J_F-4J_2\right)
\nonumber \\ 
&&\times \sum_{<k_a,k_b>}\left(\tau_{k_a}^{+}\tau_{k_b}^{-}+
\tau_{k_a}^{-}\tau_{k_b}^{+}\right)+C_1,\label{H1}
\end{eqnarray}
where the coupling is completely the same for the four distinct 
directions of pairs. In Eq.(\ref{H1}),
\begin{eqnarray}
C_1=-\frac{N}{64}\Big\{
\left(J_F-4J_2\right)^2&+&2\left(J_F-4J_1\right)^2\nonumber\\
&+&10\left(4J_2-4J_1\right)^2\Big\},
\end{eqnarray}
and the Pauli matrix $\vec{\tau}_k$, describing the two-dimensional
spin-singlet space on the $k$th tetrahedron,
is defined by using the chirality basis as,
\begin{eqnarray*}
\tau^{z}&=&|R\rangle \langle R|-|L\rangle\langle L|,\\
\tau^{+}&=&|R\rangle \langle L|,\;\;
\tau^{-}=|L\rangle \langle R|.
\end{eqnarray*}

In fact, we can show that the universality class of the effective 
Hamiltonian is $XXZ$ type by using a symmetry consideration as follows.
First, the $\pm 2\pi /3$ rotation, $C_3$ and $C_3^{-1}$, around the axis 
joining the centers of two tetrahedrons leads to the conservation of their 
total chirality. Second, the $|R\rangle$ and $|L\rangle$ states are 
mutually conjugate by the time-reversal symmetry. Therefore we can
conclude that the derived effective interaction has the $XXZ$-type
symmetry of the pseudo $\vec{\tau}$-spins and that the Ising term 
must vanish in the even-order perturbations. It should be noted that 
the second-order perturbation gives the leading terms, since
we start from the non-magnetic zeroth order states. 

In order to estimate the higher-order perturbation terms concerning $J_F$ 
within the pairwise treatment, we consider the infinite $J_F$ limit 
by considering the original spin-1 object on the sharing top vertex 
of the two tetrahedrons. In this case, the effective Hamiltonian 
obtained by the first-order perturbation is given by,
\begin{eqnarray}
h_{k_{a},k_{b}}=
\frac{1}{2.25}\left(J_2-J_1\right)
\left(
\tau_{k_a}^{+}\tau_{k_b}^{-}+\tau_{k_a}^{-}\tau_{k_b}^{+} \right)+{C_2},
\label{H2}
\end{eqnarray}
with $C_2=\left(J_2+2J_1\right)/9$, which is just 8/9
of the coefficient of the first order term of Eq. (\ref{H1}) in $J_F$.
The $\vec{\tau}$ matrixes here are defined concerning the bottom triangles,
which are related to the $\vec{\tau}$ matrix in Eq. (\ref{H1})
through the relations like
$|R\rangle =\left(
|{\uparrow_R}\rangle|\downarrow_V\rangle-
|{\downarrow_R}\rangle|\uparrow_V\rangle \right)/\sqrt{2}$,
where $|\uparrow_R\rangle$ is the doublet states with $R$ chirality
about the bottom triangle and $|\uparrow_V\rangle$ represents 
the spin state on the top vertex. It should be noted that such decompositions
cannot be extended consistently all over the lattice. Here we have shown it
to illustrate the generic form of the pairwise effective interaction
in the strong $J_F$ limit.

Comparison of Eq. (\ref{H1}) and Eq. (\ref{H2}) suggests that higher order 
effects of $J_F$ only renormalize the strength of the pairwise interaction.
Therefore, we assume that the effective Hamiltonian to describe
the low energy part of the original Heisenberg model with
next nearest neighbor ($J_1$) and third neighbor interactions ($J_2$)
is given as follows, by using the spin-1/2 pseudospin operator, 
$\vec{\tau}_k$, defined for the $k$th tetrahedron;
\begin{eqnarray}
{\cal H}_{eff}=cJ'\sum_{\langle k_a,k_b \rangle}
\left(\tau_{k_a}^{+}\tau_{k_b}^{-}+\tau_{k_a}^{-}\tau_{k_b}^{+}\right),
\label{H3}
\end{eqnarray}
where $J'=J_2-J_1$ with a positive $c$.
Since both of the upward (labeled by $k_a$) and downward ($k_b$)tetrahedrons 
form the distinct fcc lattice structure, Eq. (\ref{H3}) is the spin-1/2 
XY model on ZnS type bipartite lattice with coordination number $z=4$. 

Since the sign of the coupling constant of the XY model on a bipartite lattice
can be converted by the $\pi$-rotation around the $z$-axis concerning one of 
the two sublattices, in the following we discuss the case with the negative
interaction. For the spin-1/2 case with $N$ sites, it is exactly 
shown that the ground state is unique with $S_{tot}^z=0$ by using the
Perron-Frobenius theorem\cite{MATTIS}. As one can imagine naively, 
the state with $S_{tot}=N/2$ and $S_{tot}^z=0$ is a good zeroth order 
approximation. In fact, the Gutzwiller-type wavefunction constructed from 
this state is shown to be an extremely good variational 
state\cite{SUZUKI,OITMAA}. When applying their results for our $z=4$ case, 
the square of the long-range order (LRO) and the short-range magnetic 
correlation are given by;
\begin{eqnarray}
\langle M^{x2}\rangle /N^2 &=& 2^{15}\cdot 3^4 \cdot7^{-8}=0.4604\cdots
\label{orderpara}\\
\langle \tau_k^z\cdot \tau_{k+\delta}^z\rangle &=&-1/7,
\end{eqnarray}
where $\vec{M}=\sum_{k=1}^{N}\vec{\tau}_k$ and $\vec{\tau}_{k+\delta}$ 
means the nearest neighbor pseudospin next to the $\vec{\tau}_k$ pseudospin.
In the XY model, the Hamiltonian has the U(1) symmetry about the rotation 
around the $z$ axis in the spin space. Therefore the direction of the 
order parameter in the $\tau^x$-$\tau^y$ plane cannot be determined 
even for the infinite system. The U(1) symmetric general order parameters, 
$\vec{M}(\theta)$, are defined by the ground-state expectation value of the 
following operator;
\begin{eqnarray}
\hat{M}\left(\theta \right)=
\left(M^x\cos{\theta},M^y\sin{\theta},0\right),\label{tau}
\end{eqnarray}
where $\langle M^{x}\rangle=\langle M^{y}\rangle=M>0$. 
Concerning every local tetrahedron,
$\hat{M}(\theta)$ belongs to the $E$ representation of $T_d$.

Although the effective Hamiltonian, Eq. (\ref{H3}), has the continuous
U(1) symmetry as represented by Eq. (\ref{tau}), the original model
possesses only the finite group ($T_d$) symmetry. The U(1) symmetry
is fictitious one valid only in the lowest-order treatments
with the pseudo-spin representation. Suppose that we take account of the 
higher order interactions like biquadratic term and others, then
this spurious U(1) symmetry should be broken. In order to understand 
how this lowering of the symmetry acts, we make use of the Ginzburg-Landau 
type argument by expanding the free energy as a function of the order 
parameter $\vec{M}(\theta)$, actually by its magnitude $M$ and phase $\theta$, 
just around the second-order transition point.
The U(1) symmetry, obtained by the perturbation, corresponds to the 
second-order expansion of the free energy by $\vec{M}(\theta)$.  
This is because, on every tetrahedron, the product representation 
$E\times E$ ($=A_1+A_2+E$) contains the unique totally symmetric one ($A_1$),
which is a constant independent of $\theta$. Now we proceed to higher 
order corrections. Since under the time-reversal symmetry only $\tau^z$ 
changes the sign, while both $\tau^x$ and $\tau^y$ remain unchanged, 
the third-order expansion gives the next leading term, introducing
the anisotropy to fix $\theta$. By calculating the third order invariant,
which is obtained by reducing the product representation $E\times E\times E$,
we find that the anisotropy is proportional to $\cos{3\theta}$.
Thus we get a set of the stable states,
$\{\vec{M}(\pi),\vec{M}(\pm\pi/3)\}$ or $\{\vec{M}(0),\vec{M}(\pm2\pi/3)\}$, 
depending on the sign of the $M ^3$ term. These two sets only differ
in the overall sign and are essentially the same.

Let us illustrate the symmetry property of the order parameter 
by taking $\vec{M}(0)=\sum_{k=1}^{N}\langle \tau^x_k \rangle=
\sum_{k=1}^{N}\langle |u_k\rangle\langle u_k|-|v_k\rangle\langle v_k| \rangle$ 
as an example. The non-vanishing long-range correlation between $\tau^x_k$ and
$\tau^x_{k'}$ results in the different population of the $|u\rangle$ and 
$|v\rangle$ states on a tetrahedron. Therefore the parity symmetry concerning
the bonds vertical to the $c$-axis is broken in the ground state,
since the $|u\rangle$ and $|v\rangle$ states are characterized by the
odd and even parity concerning these bonds, respectively.
Three different values of $\theta$ in each set of order parameters
correspond to the three equivalent choice of the cubic principal axis;
$\vec{M}(2\pi/3)$ and $\vec{M}(-2\pi/3)$ characterizes the break-down of 
the parity symmetry concerning the bonds vertical to $b$ and $c$ axis, 
respectively. These three states form a domain structure in a real system.

Now we extend the present results for positive $J_F$ also and summarize 
the ground-state properties of Eq. (\ref{H3}) by taking $c$-axis as a 
principal axis ($\theta=0,\pi$). Depending on the sign of $J'=J_2-J_1$, 
the uniform ($J'<0$) or the staggered signed ($J'>0$) summation of the 
$\langle \tau^x_k\rangle$ all over the tetrahedrons defines the order 
parameter. In the classical picture, for negative $J'$ the ground state 
shows the ferromagnetic LRO of the transverse components of the 
spin chirality $|u\rangle$ ($|v\rangle$) when $\theta=0$ $(\pi)$, while 
for positive $J'$ the antiferromagnetic LRO of the $|u\rangle$ and 
$|v\rangle$ states. For a local tetrahedron problem the compressed and 
elongated $Q_v$ mode along the $c$-axis stabilize the $|u\rangle$ and 
$|v\rangle$ states, respectively\cite{YAMASHITA}. 
Therefore, for the ferromagnetic $J'$ model the order parameter, 
$\vec{M}(0)$ $(\vec{M}(\pi))$, induces the uniform $Q_v$ lattice distortion 
with $c/a<1$ ($>1$). On the other hand, for the AF $J'$ the staggered $Q_v$ 
distortion is induced.
Accordingly, under the assumption that the inter-tetrahedron interactions
are more important than the local magneto-elastic couplings,
the inter-tetrahedron couplings ($J_1$ and $J_2$) lift the many-fold
degeneracy of the spin-singlet manifold by the second-order phase transition.
However, introduction of the magneto-elastic coupling, even for
a weak case, will induce a small first order structural distortion,
leading eventually to a weak first order phase transition.

In conclusion, we have studied the ground-state properties of the spin-1 
pyrochlore antiferromagnet by using the tetrahedron-unit decomposition 
of the pyrochlore lattice. In this approach, we assume that the spin-singlet 
manifold spanned by the VBS-type wavefunction well-describes the low-energy 
physics and the essential point is how the degeneracy is lifted.
We have investigated the case where this degeneracy is lifted by the
inter-tetrahedron interactions caused by the next nearest neighbor
and the third neighbor interaction, which produce the the pairwise effective 
Hamiltonian of the $XY$ type between the pseudospin operators describing 
the tetrahedron singlets. To break the spurious U(1) symmetry of the 
effective model, we have considered the higher order anisotropic term based
on the symmetry property ($E$) of the order parameter. It has turned out that
the parity-broken ground state emerges through the second-order phase 
transition.

Lastly we briefly comment on the relation between the present results and 
the numerical results for the pure spin-1 Heisenberg model\cite{KOGA}. 
The singlet ground states of a single spin-1 tetrahedron problem are 
three dimensional with $A_1+E$ irreducible representations.
When we define these spin-singlet states,
with $\vec{S}_{1}+\vec{S}_{2}=\vec{S}_{3}+\vec{S}_{4}=0$, 1, and 2,
by $|a\rangle, |b\rangle$, and $|c\rangle$, respectively,
the orthonormal basis of the $E$ representation are given by 
$\{|U\rangle,|V\rangle\}=\{|b\rangle,(-2|a\rangle+\sqrt{5}|c\rangle)/3\}$.
Considering the fact that the symmetrization on every vertex does not affect 
the local symmetry properties, the parity-broken symmetry of the ground state, 
obtained by using the decomposite-spin representation, may correspond to 
the $|U\rangle$ and $|V\rangle$ states in the spin-1 picture.
Since Koga $et.\,al$ suggested the possibility of the new spin-gap
ground state characterized by the $|U\rangle$ state around
the isotropic Heisenberg limit \cite{KOGA}, our result seems to be consistent with 
their result.

We are grateful to H. Tsunetsugu, A. Koga, and  N. Kawakami 
for useful comments and discussions. 
Y.Y. is supported by the Japan Society for the Promotion of Science.

\end{multicols}
\end{document}